\documentclass[prd,superscriptaddress,amsmath,twocolumn,nofootinbib]{revtex4-2}

\pdfoutput=1
\usepackage{times}

\usepackage{graphicx}
\usepackage{xcolor}

\usepackage{array}
\usepackage{booktabs}
\usepackage{enumitem}
\usepackage{epsf,latexsym,bbm,euscript}

\usepackage{mathrsfs,mathtools,amsmath,amssymb}
\usepackage{tensor}

\makeatletter
\newcommand*{\defeq}{\mathrel{\rlap{%
			\raisebox{0.3ex}{$\m@th\cdot$}}%
		\raisebox{-0.3ex}{$\m@th\cdot$}}%
	=}
\newcommand*{\eqdef}{=\mathrel{\rlap{%
			\raisebox{0.3ex}{$\m@th\cdot$}}%
		\raisebox{-0.3ex}{$\m@th\cdot$}}%
}
\makeatother

\newcommand{\RNum}[1]{\uppercase\expandafter{\romannumeral #1\relax}}

\usepackage{scalerel}
\usepackage{tikz}
\usetikzlibrary{svg.path}
\definecolor{orcidlogocol}{HTML}{A6CE39}
\tikzset{
	orcidlogo/.pic={
		\fill[orcidlogocol] svg{M256,128c0,70.7-57.3,128-128,128C57.3,256,0,198.7,0,128C0,57.3,57.3,0,128,0C198.7,0,256,57.3,256,128z};
		\fill[white] svg{M86.3,186.2H70.9V79.1h15.4v48.4V186.2z}
		svg{M108.9,79.1h41.6c39.6,0,57,28.3,57,53.6c0,27.5-21.5,53.6-56.8,53.6h-41.8V79.1z M124.3,172.4h24.5c34.9,0,42.9-26.5,42.9-39.7c0-21.5-13.7-39.7-43.7-39.7h-23.7V172.4z}
		svg{M88.7,56.8c0,5.5-4.5,10.1-10.1,10.1c-5.6,0-10.1-4.6-10.1-10.1c0-5.6,4.5-10.1,10.1-10.1C84.2,46.7,88.7,51.3,88.7,56.8z};
	}
}
\newcommand\orcidlink[1]{\href{https://orcid.org/#1}{\mbox{\scalerel*{
				\begin{tikzpicture}[yscale=-1,transform shape]
					\pic{orcidlogo};
				\end{tikzpicture}
			}{X}}}}

\usepackage{url,hyperref}
\hypersetup{colorlinks,linkcolor={blue!55!black},citecolor={red!50!black},urlcolor={blue!45!black},breaklinks=true}

\begin{document}

\title{Quantum tunneling from Schwarzschild black hole in non-commutative gauge theory of gravity}

\author{Abdellah Touati\orcidlink{0000-0003-4478-2529}}
\email{abdellah.touati@univ-batna.dz, (touati.abph@gmail.com)}
\affiliation{Department of Physics, Faculty of Matter Sciences, University of Batna-1, Batna 05000, Algeria}

\author{Slimane Zaim}
\email{zaim69slimane@yahoo.com}
\affiliation{Department of Physics, Faculty of Matter Sciences, University of Batna-1, Batna 05000, Algeria}

\begin{abstract}
In this letter, we present the first study of Hawking radiation as a tunneling process within the framework of non-commutative (NC) gauge theory of gravity. First, we reconstruct the non-commutative Schwarzschild black hole (NC SBH) within the gauge theory of gravity, employing the Seiberg-Witten (SW) map and the star product. Then, we compute the emission spectrum of outgoing massless particles using the quantum tunneling mechanism. In the first scenario, we calculate the tunneling rate of massless particles crossing the event horizon of the NC SBH with lower frequencies. Our results reveal pure thermal radiation. Notably, we find that the Hawking temperature remains consistent in both the classical thermodynamics and the quantum tunneling approach, suggesting equivalence between these two approaches in NC spacetime. However, in the case of massless particle emission with higher frequencies, we account for energy conservation resulting in the tunneling rate to deviate from pure thermal radiation. This tunneling rate remains consistent with an underlying unitary quantum theory. We establish a relationship between this deviation and the change in the black hole entropy, revealing a logarithmic correction to the entropy within this geometry. Furthermore, we demonstrate that non-commutativity enhances the correlations between two successively emitted particles. Additionally, we determine the NC density number of particle emission and conclude by discussing the implications of our findings.
\end{abstract}

\maketitle

\section{Introduction} \label{sec:introduction}

In 1975, S. Hawking published his seminal work \cite{hawking1}, which established a pivotal connection between quantum processes and gravity, serving as the foundational bridge to the quantum theory of gravity. In this groundbreaking paper, Hawking demonstrated that black holes can emit pure thermal radiation, effectively undergoing evaporation \cite{hawking1,hawking3}. His calculations were rooted in quantum field theory applied within the context of curved spacetime near the event horizon. This remarkable phenomenon, known as Hawking radiation, opened the door to the exploration of black hole thermodynamics, becoming an indispensable tool for investigating and comprehending quantum effects near the event horizon in the realm of black hole physics \cite{hawking3,harms,vaz,haranas2,hansen1,chen1,hansen,jawad1}.

More recently, a novel approach was initiated by Kraus and Wilczek \cite{kraus1} and further developed by Parikh and Wilczek \cite{tunn1,tunn3,tunn2}. This approach characterizes Hawking radiation as a tunneling process within a semi-classical framework. This method has proven to be invaluable for studying Hawking radiation in a wide range of black holes characterized by static spherical symmetry \cite{tunn10,tunn12,tunn11,tunn13,tunn14,tunn15,tunn16,tunn17,tunn5,tunn18,tunn6,tunn8,tunn9,tunn7,tunn4}.

In this letter, our objective is to expand upon this method for calculating the tunneling rate from deformed SBH in the NC gauge theory. This theory is motivated by string theory, as discussed in \cite{seiberg1}. The central concept of the NC theory involves the quantization of gravity through the quantization of spacetime. This is achieved by enforcing a commutation relation between the coordinates themselves
\begin{equation}
[x^{\mu},x^{\nu}]=i\Theta^{\mu\nu}\,,
\end{equation}
where $\Theta^{\mu\nu}$ is an anti-symmetric real matrix. This commutation relation modifies the ordinary product of two arbitrary functions $f(x)$ and $g(x)$, defined over this spacetime, leading to the introduction of a new product known as the star product (or Moyal product) denoted as ``$*$'', defined as:
\begin{equation}\label{eqt2.25}
(f*g)(x)=f(x)e^{\frac{i}{2}\Theta^{\mu\nu}\overleftarrow{\partial_{\mu}}\overrightarrow{\partial_{\nu}}}g(x)\,.
\end{equation}

It is worth noting that previous studies have explored the tunneling process from the NC black hole using the ``coordinate coherent states approach'' \cite{nctunn3,nctunn6,nctunn2,nctunn1,nctunn5,nctunn7,nctunn4}. Our research is inspired by our recent paper \cite{abdellah2}, in which we investigated the thermodynamic properties of SBH in the NC gauge gravity. In that study, we applied classical black hole thermodynamics, utilizing surface gravity to derive the deformed Hawking temperature and the black hole area for calculating entropy and other properties. Our investigation led to two significant findings: the identification of a new fundamental length at the Planck scale and a novel scenario for black hole evaporation.

In the current work, we focus on examining various thermodynamic properties of NC SBH using a quantum process known as quantum tunneling. This analysis involves the application of the gauge theory of gravity, the star product, and the SW maps \cite{seiberg1}.

In this study, we derive the NC SBH in the gauge theory of gravity. We obtain corrections to the tunneling rate from NC SBH in two cases: one involving lower frequencies, which exhibit pure thermal radiation and align with thermodynamic processes, and another involving conservation of energy. In the latter case, the results deviate from thermal radiation but remain consistent with an underlying unitary theory. Additionally, we investigate the Hawking temperature within this geometry and derive a logarithmic correction to the NC black hole's entropy. We also explore the correlation between successive emissions of particles. Finally, we calculate the NC correction to the density of particles created, and our results demonstrate that non-commutativity plays a role similar to mass. Increasing the strength of the gravitational field of the deformed SBH leads to a decrease in the density of particles tunneling from this black hole.

\section{Non-commutative Schwarzschild black hole}\label{sec:NCSBH}

We briefly review the formalism of the NC gauge theory of gravity \cite{cham1,chai1,chai2,mukhe1}, which generalizes the classical theory based on the de Sitter group as a local symmetry \cite{zet1,zet2}. We employ the tetrad formalism, along with both the star ($\ast$-) product and the SW map, to construct an NC gauge theory for a Schwarzschild metric. The NC corrections to the commutative tetrad fields $\hat{e}^{a}_{\mu}(x,\Theta)$ are obtained using the perturbation form for the SW map, described as a power series in $\Theta$ up to second order \cite{cham1}
\begin{equation}
\hat{e}^{a}_{\mu}(x,\Theta)=e^{a}_{\mu}(x)-i\Theta^{\nu\rho}e^{a}_{\mu\nu\rho}(x)+\Theta^{\nu\rho}\Theta^{\lambda\tau}e^{a}_{\mu\nu\rho\lambda\tau}(x)+\mathcal{O}(\Theta^{3})\label{eqt2.26}\,,
\end{equation}
where
\begin{equation}
e^{a}_{\mu\nu\rho}=\frac{1}{4}[\omega^{ac}_{\nu}\partial_{\rho}e^{d}_{\mu}+(\partial_{\rho}\omega^{ac}_{\mu}+R^{ac}_{\rho\mu})e^{d}_{\nu}]\eta_{cd}\,.
\end{equation}
Then we have
\small
\begin{widetext}
\begin{align}
\hat{e}_\mu^a&=\frac{\Theta^{\nu\rho}\Theta^{\lambda\tau}}{32}\left[2\{R_{\tau\nu},R_{\mu\rho}\}^{ab}e^{c}_{\lambda}-\omega^{ab}_{\lambda}(D_{\rho}R_{\tau\nu}^{cd}+\partial_{\rho}R_{\tau\nu}^{cd})e^{m}_{\nu}\eta_{dm}-\{\omega_{\nu},(D_{\rho}R_{\tau\nu}+\partial_{\rho}R_{\tau\nu})\}^{ab}e^{c}_{\lambda}-\partial_{\tau}\{\omega_{\nu},(\partial_{\rho}\omega_{\mu}+R_{\rho\mu})\}^{ab}e^{c}_{\lambda}\right.\notag\\
&\left.-\omega^{ab}_{\lambda}\left(\omega^{cd}_{\nu}\partial_{\rho}e^{m}_{\mu}+\left(\partial_{\rho}\omega_{\mu}^{cd}+R_{\rho\mu}^{cd}\right)e^{m}_{\nu}\right)\eta_{dm}+2\partial_{\nu}\omega_{\lambda}^{ab}\partial_{\rho}\partial_{\tau}e^{c}_{\mu}-2\partial_{\rho}\left(\partial_{\tau}\omega_{\mu}^{ab}+R_{\tau\mu}^{ab}\right)\partial_{\nu}e^{c}_{\lambda}-\{\omega_{\nu},(\partial_{\rho}\omega_{\lambda}+R_{\rho\lambda})\}^{ab}\partial_{\tau}e^{c}_{\mu}\right.\notag\\
&\left.-\left(\partial_{\tau}\omega_{\mu}+R_{\tau\mu}\right)\left(\omega^{cd}_{\nu}\partial_{\rho}e^{m}_{\lambda}+\left((\partial_{\rho}\omega_{\lambda}+R_{\rho\lambda})\right)e^{m}_{\nu}\right)\eta_{dm}\right]\eta_{cb}+\mathcal{O}\left( \Theta^{3}\right),\label{eq:SWM}
\end{align}
\end{widetext}
\normalsize
where $\hat{e}_{a}^{\mu }$ and $\omega^{ab}_{\mu}$ are the commutative tetrad field and spin connection, and
\begin{align}
\{\alpha,\beta\}^{ab}&=\left(\alpha^{ac}\beta^{db}+\beta^{ac}\alpha^{db}\right)\eta_{cd},\\ [\alpha,\beta]^{ab}&=\left(\alpha^{ac}\beta^{db}-\beta^{ac}\alpha^{db}\right)\eta_{cd}\\
D_{\mu}R_{\rho\sigma}^{ab}&=\partial_{\mu}R^{ab}_{\rho\sigma}+\left(\omega_{\mu}^{ac}R^{db}_{\rho\sigma}+\omega_{\mu}^{bc}R^{da}_{\rho\sigma}\right)
\end{align}
where $\hat{e}_{a}^{\mu }$ is the inverse of the vierbein $\hat{e}_{\mu }^{a}$ defined as
\begin{equation}
\hat{e}_{\mu }^{b} \hat{e}_{a}^{\mu }=\delta _{a}^{b},\quad \hat{e}_{\mu }^{a} \hat{e}_{a}^{\nu }=\delta _{\mu }^{\nu }\,.
\end{equation}

For the NC metric $\hat{g}_{\mu\nu}$, we use the expression found in Ref. \cite{chai1}
\begin{equation}\label{eq:metric}
\hat{g}_{\mu \nu }=\frac{1}{2}(\hat{e}_{\mu }^{b}\ast \hat{e}_{\nu b}+\hat{e}_{\nu }^{b}\ast \hat{e}_{\mu b})\,.
\end{equation}
To compute the deformed metric $\hat{g}_{\mu \nu}$, we choose the following NC anti-symmetric matrix $\Theta^{\mu\nu}$
\begin{equation}
\Theta^{\mu\nu}=\left(\begin{matrix}
	0	& 0 & 0 & 0 \\
	0	& 0 & \Theta & 0 \\
	0	& -\Theta & 0 & 0 \\
	0	& 0 & 0 & 0
\end{matrix}
\right), \qquad \mu,\nu=0,1,2,3\label{eqt2.34}\,.
\end{equation}
Following the same steps outlined in Ref. \cite{abdellah1}, we select the following general tetrad field
\begin{align}
\underline{e}_{\mu }^{0}&=\left(\begin{array}{cccc}\left(1-\frac{2 m}{r}\right)^{\frac{1}{2}}, & 0, & 0, & 0\end{array}\right),\notag \\
\underline{e}_{\mu }^{1}&=\left(\begin{array}{cccc}0, & \left(1-\frac{2 m}{r}\right)^{-\frac{1}{2}}\sin\theta \cos\phi, & r \cos\theta \cos\phi, & -r \sin\theta \sin\phi\end{array}\right), \notag\\
\underline{e}_{\mu }^{2}&=\left(\begin{array}{cccc}0, & \left(1-\frac{2 m}{r}\right)^{-\frac{1}{2}}\sin\theta \sin\phi, & r \cos\theta \sin\phi, & r \sin\theta \cos\phi\end{array}\right), \notag\\
\underline{e}_{\mu }^{3}&=\left(\begin{array}{cccc}0, & \left(1-\frac{2 m}{r}\right)^{-\frac{1}{2}}\cos\theta, & -r \sin\theta, & 0\end{array}\right). \label{eq:tetrad}
\end{align}

We calculate the components of the deformed metric $\hat{g}_{\mu\nu}$ for the Schwarzschild black hole using Eqs. \eqref{eq:SWM} together with Eq. \eqref{eq:tetrad}. By employing the definition \eqref{eq:metric}, we obtain the non-zero components of the NC metric $\hat{g}_{\mu \nu }$ at leading order in $\Theta$
\small
\begin{widetext}
\begin{align}
-\hat{g}_{00}&=\left(1-\frac{2 m}{r}\right)+\Theta^{2}\left\{\frac{m\left(88m^2+mr\left(-77+15\sqrt{1-\frac{2m}{r}}\right) -8r^2\left(-2+\sqrt{1-\frac{2m}{r}}\right)\right)}{16r^4(-2m+r)}\right\}+\mathcal{O}(\Theta^4),\label{eq:13}\\
\hat{g}_{11}&=\left(1-\frac{2 m}{r}\right)^{-1}+\Theta^{2}\left\{\frac{m\left(12m^2+mr\left(-14+\sqrt{1-\frac{2m}{r}}\right)-r^2\left(5+\sqrt{1-\frac{2m}{r}}\right)\right)}{8r^2(2m-r)^3}\right\}+\mathcal{O}(\Theta^{4}),\label{eq:14}\\
\hat{g}_{22}&=r^{2}+\Theta^2\left\{\frac{-18m^2\left(-6+\sqrt{1-\frac{2m}{r}}\right)r+m\left(-57+29\sqrt{1-\frac{2m}{r}}\right)r^22\left(5-3\sqrt{1-\frac{2m}{r}}\right) r^3-68m^3}{16r(-2m+r)^2}\right\}+\mathcal{O}(\Theta^4),\label{eq:15}\\
\hat{g}_{33}&=r^{2}sin^2(\theta)+\Theta^{2}\left\{\frac{m^2\left(50-6\sqrt{1-\frac{2m}{r}}\right)r+m\left(-43+23\sqrt{1-\frac{2m}{r}}\right)r^2+2\left(5-3\sqrt{1-\frac{2m}{r}}\right)r^3}{32r(r-2m)^2}\notag\right.\\
&\left.+\frac{-8m^3+(2m-r)\left(4 m^2 + 3 m (-1 + \sqrt{1-\frac{2m}{r}}) r - 2 (1 + \sqrt{1-\frac{2m}{r}}) r^2\right)cos(2\theta)}{32r(r-2m)^2}\right\}+\mathcal{O}(\Theta^4).\label{eq:16}
\end{align}
\end{widetext}
\normalsize
It is evident that as $\Theta\rightarrow0$, we recover the commutative Schwarzschild solution. The NC line element, based on the metric above, is given by
\small
\begin{equation}\label{eq:line-element}
	d\hat{s}^{2}=-\hat{g}_{00}(r,\Theta)dt^{2}+\hat{g}_{11}(r,\Theta)dr^{2}+\hat{g}_{22}(r,\Theta)d\theta^{2}+\hat{g}_{33}(r,\Theta) d\phi^{2},
\end{equation}
\normalsize
For this black hole, we determine the NC event horizon by identifying the point where the NC metric \eqref{eq:13} satisfies the condition $\frac{1}{\hat{g}_{11}}=0$. The solution yields the NC event horizon of the SBH \cite{abdellah2}
\begin{equation}\label{eq:horizon}
	r_{h}^{NC}=r_{h}\left[1+\frac{3}{8}\left(\frac{\Theta}{r_h}\right)^2\right]
\end{equation}
where $r_{h}=2m$ is the event horizon in the commutative case when $\Theta =0$. The influence of non-commutativity is small, which is reasonable to expect, as it becomes negligible at large distances.

\section{Quantum tunneling in NC spacetime} \label{sec:QTNC}

As per the authors in Refs. \cite{tunn1,tunn2,tunn3}, it is imperative to employ stationary coordinates in the description of the tunneling process. These stationary coordinates are in contrast to Schwarzschild coordinates, which exhibit no singularity at the event horizon, thereby guaranteeing the conservation of energy in the radiation spectrum derivative. To facilitate this, we represent the NC SBH metric in the Painlevé-Gullstrand form \cite{tunn1,tunn2,tunn3,tunn4}. The NC line element \eqref{eq:line-element} for radial motion becomes
\begin{align}\label{eq:PG-line}
d\hat{s}^{2}&=-\hat{g}_{00}(r,\Theta)dt^{2}+2\hat{h}(r,\Theta)dtdr+dr^{2}+\hat{g}_{22}(r,\Theta)d\theta^{2}\notag\\
&+\hat{g}_{33}(r,\Theta) d\phi^{2}\,,
\end{align}
where 
\begin{equation}
\hat{h}(r,\Theta)=\sqrt{\hat{g}_{00}(r,\Theta)(\hat{g}_{11}(r,\Theta)^{-1}-1)}
\end{equation}

In the semi-classical tunneling of particles, the tunneling rate is related to the imaginary part of the action \cite{tunn1,tunn2,tunn3}
\begin{equation}
\Gamma\sim e^{-\text{Im}\hat{S}}\,.
\end{equation}
Here, $\hat{S}$ is the tunneling action of particles in the NC curved spacetime, which is defined as
\begin{equation}
\hat{S}=\int p_{\mu}dx^{\mu}\,,
\end{equation}
where $p_{\mu}=\hat{g}_{\mu\nu}\frac{dx^{\nu}}{d\tau}$ is the conjugate momentum, and $\tau$ is the affine parameter. We are interested in the tunneling of a massless particle with radial motion (in this case, the angular momentum is zero) and in the equatorial plane $\theta=\pi/2$. The imaginary part of the action is reduced to
\begin{align}\label{eq:ims}
\text{Im}\hat{S}&=\text{Im}\int(p_{t}dt+p_rdr)=\text{Im}\int_{r_i}^{r_f} \int_{0}^{p_r}dp'_r dr.
\end{align}

Observing that the first term of the integration is real and, therefore, does not contribute to the calculation of the imaginary part of the action, we turn to the Hamiltonian equation. In this context, the system's Hamiltonian is represented as $H=m-\omega'$. This allows us to write
\begin{equation}
\dot{r}=\frac{dH}{dp_r}=\frac{d(m-\omega')}{dp_r},
\end{equation}
where $\dot{r}=\frac{dr}{dt}$. Incorporating these considerations into Eq. \eqref{eq:ims} we derive the expression
\begin{align}\label{eq:ims2}
\text{Im}\hat{S}=\text{Im} \int_{m}^{m-\omega}\int_{r_i}^{r_f} \frac{d(m-\omega)}{\dot{r}}dr.
\end{align}

We assume that the particle undergoing tunneling across the event horizon follows a radial trajectory within the equatorial plane, defined by $\theta=\frac{\pi}{2}$, hence $d\theta=0$. Consequently, the radial null geodesic can be determined by utilizing the following expression $\hat{g}_{\mu\nu}U^\mu U^\nu=0$
\begin{align}\label{eq:radial}
	\frac{dr}{dt}=-\hat{g}_{01}+\sqrt{\hat{g}_{01}^2+\hat{g}_{00}}.
\end{align}
Inserting the above equation into Eq. \eqref{eq:ims2}, we obtain
\begin{align}\label{eq:ims3}
\text{Im}\hat{S}=\text{Im} \int_{m}^{m-\omega}\int_{r_i}^{r_f} \frac{d(m-\omega)}{-\hat{g}_{01}+\sqrt{\hat{g}_{01}^2+\hat{g}_{00}}}dr.
\end{align}
where $\hat{g}_{01}=\hat{h}$ and $dH=d(m-\omega)=-d\omega$. After rearranging our expression, we find
\begin{align}\label{eq:ims4}
\text{Im}\hat{S}=\text{Im} \int_{0}^{\omega}(-d\omega)\int_{r_i}^{r_f} \frac{dr}{\sqrt{\hat{g}_{00}\hat{g}_{11}}\left[1-\sqrt{1-\frac{1}{\hat{g}_{11}}}\right]}.
\end{align}
This expression exhibits a singular pole at the NC event horizon $r=r^{NC}_h$. To compute this integral, we employ a contour deformation technique around the pole, making use of the residue theorem. To ensure that the Boltzmann factor is conserved (in the approximation of lower frequency $\omega\ll m$), $\Gamma\sim e^{-2\text{Im}S}\sim e^{-\beta \omega}$, we follow the same steps to evaluate the integral over $\omega$ as in Refs. \cite{tunn9,tunn7}:
\begin{align}\label{eq:ims5}
\text{Im}\hat{S}&=\pi \int_{0}^{\omega}\left(4m+\frac{3\Theta^2}{2m}\right)d\omega=\pi\omega\left(4m+\frac{3\Theta^2}{2m}\right).
\end{align}
The tunneling rate from the NC SBH is therefore given by
\begin{align}\label{eq:tunneling1}
\hat{\Gamma}\sim \exp\left[-2\pi\omega\left(4m+\frac{3\Theta^2}{2m}\right)\right].
\end{align}
In the limit $\Theta \rightarrow 0$, the commutative expression is recovered \cite{tunn1,tunn2}.

\begin{figure}[h]
	\centering
	\includegraphics[width=0.45\textwidth]{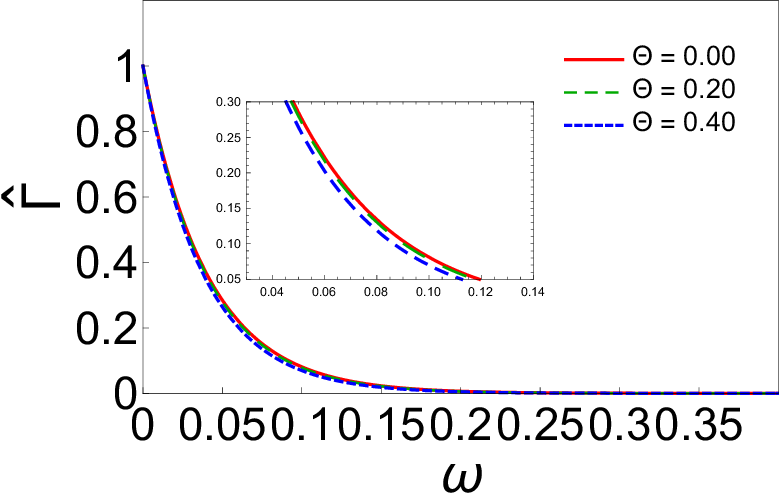}
	\caption{The behavior of tunneling rate from NC SBH as a function of the emitting particle energy $\omega$, with $m=1$.}
	\label{fig1}
\end{figure}

In Figure \ref{fig1}, we illustrate the impact of non-commutativity on the tunneling rate of particles from the NC SBH. It is evident that non-commutativity has a minor influence on the tunneling rate for emitted particles with frequencies in the range $\omega\in [0.04,0.15]$. In this range, an increase in the non-commutativity parameter $\Theta$ leads to a decrease in $\hat{\Gamma}$. For other values of $\omega$ or when $\Theta$ is small, the effect of non-commutativity becomes negligible.
\subsection{NC correction to the Hawking temperature}\label{subsec:NCT}

It is important to highlight that, according to Planck's law, particles with frequency $\omega$ are expected to be emitted at a rate described by $\Gamma\sim \exp\left[-\omega/T\right]$. Consequently, the black hole temperature, at leading order in $\Theta$, as indicated by eq. \eqref{eq:tunneling1}, can be expressed as follows
\begin{align}\label{eq:tempertature}
\hat{T}_H&=\frac{1}{2\pi\left(4m+\frac{3\Theta^2}{2m}\right)}=T_H\left(1-\frac{3\Theta^2}{2r_h^2}\right).
\end{align}
In the scenario where $\Theta=0$, the NC Hawking temperature reduces to the commutative temperature $T_H=\frac{1}{4\pi r_h}$.
\begin{figure}[h]
\centering
\includegraphics[width=0.45\textwidth]{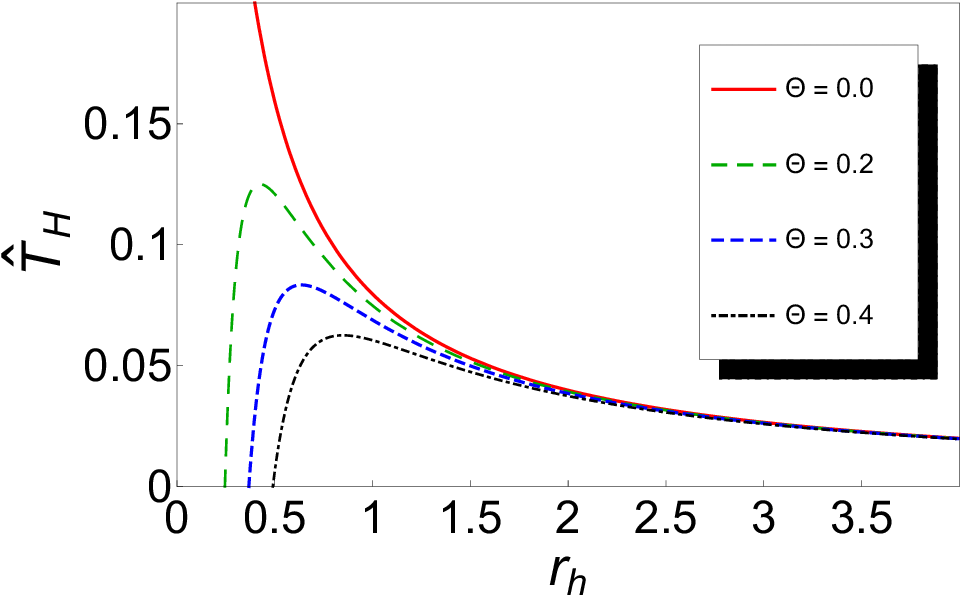}
\caption{Behavior of the Hawking temperature as a function of the black hole mass $r_h$.}
	\label{fig2}
\end{figure}

In Fig. \ref{fig2}, we present the behavior of the Hawking temperature as a function of the black hole event horizon $r_h$. Notably, the non-commutativity of spacetime mitigates the divergence of the Hawking temperature, leading the black hole to attain a new maximum temperature, $\hat{T}^{max}_H\approx\frac{0.025}{\Theta}$, at the critical size $r_h^{crit}\approx 2.121\,\Theta$. Beyond this critical point, the temperature decreases and eventually reaches zero at $r_h^{min}=1.225\Theta$, signifying the cessation of the black hole's evaporation process.

It is crucial to emphasize that the expression \eqref{eq:tempertature} aligns with the results obtained in our previous work, employing surface gravity, as discussed in detail in \cite{abdellah2}. This finding indicates the preservation of equivalence between the quantum tunneling process of black hole evaporation and the thermodynamic approach, despite the NC nature of spacetime.

The estimation of the NC parameter $\Theta$ can be derived from the point of maximum temperature, where the thermal energy is given (in natural units $\hbar=k_B=c=1$) by $E_{th}=\hat{T}^{max}_H \approx \frac{0.025}{\Theta}$ at the critical radius $r_h^{crit}\approx 2m \approx 2.121\,\Theta$. The total mass of the black hole is calculated as $M=\frac{1}{2}r_h^{crit}\frac{1}{G}\approx 1.0605\, \Theta\, M_{Planck}^2$, where we utilize the reduced Planck mass $M_{Planck}=2.435\times 10^{18}$ GeV. Consequently, the NC parameter can be estimated as follows:
\begin{equation}\label{eq:32}
\Theta \approx 1.23585\times 10^{-35}m\sim l_{Planck}.
\end{equation}

This result reveals that the NC parameter is situated at the Planck scale, corroborating results obtained through gravitational-wave experimental data \cite{ThetaGW} and our previous work on the thermodynamics of NC SBH \cite{abdellah2}. Notably, some papers have established a bound on the NC parameter via the study of black hole thermodynamics, e.g., \cite{piero1, nicolini2, alavi, bound1}, with expectations that it should be around $\sqrt{\Theta}\sim 10^{-1}.l_p$. It is worth mentioning that in our earlier research, where we examined the fourth classical test within this geometry \cite{abdellah1, abdellah3}, we established that the lower bound of the NC parameter $\Theta$ falls within the range of ($10^{-31}m-10^{-34}m$). However, in the context of thermodynamics \cite{abdellah2} (as also seen in eq. \eqref{eq:32}), we derived a new estimate of $\Theta$ at the Planck scale, i.e., $10^{-35}m$. This suggests that estimating $\Theta$ through thermal phenomena provides a more accurate determination and that is due to the direct relation of this phenomenon to the event horizon of the black hole, where this last is affected by the non-commutativity, this direct relation provides a better estimation of the NC parameter, and that means that the closer we are to the event horizon or in the presence of a strong gravitational field, the more quantum effects emerge at a large scale. In accordance with this result and in conjunction with our prior works \cite{abdellah1, abdellah2, abdellah3}, as well as references in the literature such as \cite{piero1, nicolini2, alavi, bound1, ThetaGW}, it is evident that the NC property of spacetime manifests itself in proximity to the Planck scale.

\subsection{Logarithmic corrections to the entropy in NC spacetime}\label{subsec:NCLS}

In this subsection, we will extract the entropy correction arising from the tunneling process, particularly when dealing with a large frequency $\omega$, while ensuring the conservation of energy. To facilitate this, we will employ the definition \eqref{eq:ims3} and integrate the relation \eqref{eq:ims5} over the quantity $d(\hat{m}-\omega)$. Here, we note that $\hat{m} = r^{NC}_h/2$ represents the NC correction to the black hole mass, and in the NC case, the integration element becomes $dH = d(m-\omega) \rightarrow d\hat{H} = \left(1 - \frac{3\Theta^2}{32(m-\omega)^2}\right)d(m-\omega)$. We integrate this relation from the initial state $m$ to the final state $m-\omega$ while conserving energy. According to Refs. \cite{tunn1, tunn2}, the mass of the black hole in \eqref{eq:ims5} is replaced by $m-\omega$. This leads to the following expression:
\begin{align}\label{eq:ims6}
\text{Im}\hat{S}=&-\pi \int_{m}^{m-\omega}\left(4(m-\omega)+\frac{3\Theta^2}{2(m-\omega)}\right)d(m-\omega)\notag\\
&+\pi\int_{m}^{m-\omega}\left(\frac{3\Theta^2}{8(m-\omega)}\right)d(m-\omega)+\mathcal{O}(\Theta^4),\notag\\
=&-\pi\left(2(m-\omega)^2+\frac{9\Theta^2}{16}\ln\left(4\pi(m-\omega)^2\right)-2(m)^2\right.\notag\\
&\left.-\frac{9\Theta^2}{16}\ln\left(4\pi m^2\right)\right).
\end{align}
This leads us to the usual relationship between the tunneling rate and entropy \cite{tunn1, tunn2, tunn3}
\begin{align}\label{eq:tunneling2}
\hat{\Gamma}\sim e^{-2Im\hat{S}}=e^{\Delta\hat{S}_{BH}}.
\end{align}
where $\Delta\hat{S}_{BH}= \hat{S}_{BH}(m-\omega)-\hat{S}_{BH}(m)$ is the difference in Bekenstein-Hawking entropy for the NC SBH. It is given by
\begin{align}\label{eq:entropy1}
\hat{S}_{BH}=4\pi m^2+\frac{9\pi\Theta^2}{8}\ln(4\pi m^2).
\end{align}

In the NC spacetime, we observe that the relationship between the black hole area and entropy is not conserved, and the NC correction term introduces a logarithmic correction to the entropy. This result is in contrast to our previous work \cite{abdellah2}. Interestingly, this finding aligns with similar outcomes in string theory and loop quantum gravity (at leading order in $\alpha$) \cite{tunn5}
\begin{align}\label{eq:entropy2}
\hat{S}_{BH}=\frac{A}{4}+\alpha\, \ln(A)\,,
\end{align}
where $A=4\pi m^2$ is the commutative area of the SBH, and $\alpha$ is a constant, which varies depending on the specific theory.

In Quantum Loop Gravity (QLG), $\alpha$ assumes a negative value, $\alpha = -\frac{1}{2}$ \cite{QLG1, QLG2}, while in String theory (ST), $\alpha$ is a 4-D central charge that depends on the number of fields and can take both negative and positive values \cite{ST3}. Furthermore, this logarithmic correction is also observed in the generalized uncertainty principle (GUP) and modified dispersion relation (MDR) \cite{nozari3, nozari4}. In Ref. \cite{nozari3}, a significant connection is established between the prefactor of the logarithmic correction and the spacetime dimension.

In our work using the NC gauge theory of gravity, it is worth noting that $\alpha$ consistently takes a positive value, closely related to the NC parameter by $\alpha=\frac{9\pi\Theta^2}{8}$. By using our result for the NC parameter values \eqref{eq:32}, we can relate this constant to the Planck length as $\alpha\sim \frac{9\pi}{4}l_{Planck}^2$. This indicates that in the NC gauge theory, the coefficient of the logarithmic correction to the entropy represents a quantum area at the Planck level, suggesting that spacetime is quantized at the Planck scale due to non-commutativity.

On another note, the NC correction to the tunneling rate \eqref{eq:tunneling2} can be expressed as \cite{tunn5}, according to Eq. \eqref{eq:ims6}, the difference in entropy its given by

\begin{align}\label{eq:deltaentropy}
	\Delta\hat{S}&=\pi\left(4(m-\omega)^2+\frac{9\Theta^2}{8}\ln\left(4\pi(m-\omega)^2\right)-4(m)^2\right.\notag\\
	&\left.-\frac{9\Theta^2}{8}\ln\left(4\pi m^2\right)\right),\notag\\
	&=-8\pi m\omega\left(1-\frac{\omega}{2m}\right)+\frac{9\pi\Theta^2}{8}\ln\left(1-\frac{\omega}{m}\right)^2.
\end{align}
Substituting this equation inside Eq. \eqref{eq:tunneling2}, we get
\begin{align}\label{eq:tunuling2'}
	\hat{\Gamma}&=e^{\Delta\hat{S}_{BH}}=e^{-\pi\left(-8m\omega\left(1-\frac{\omega}{2m}\right)+\frac{9\Theta^2}{4}\ln\left(1-\frac{\omega}{m}\right)^2\right)}\notag\\
	&=\left(1-\frac{\omega}{m}\right)^{\frac{9\pi}{4}\Theta^2}e^{-8\pi m\omega\left(1-\frac{\omega}{2m}\right)}.
\end{align}
By using our results in Eq. \eqref{eq:32}, $\Theta\simeq l_{Planck}$. The above equation can be written in this form
\begin{align}\label{eq:tunneling3}
\hat{\Gamma}=e^{\Delta\hat{S}_{BH}}=\left(1-\frac{\omega}{m}\right)^{\frac{9\pi}{4}l_{Planck}^2}e^{-8\pi m\omega\left(1-\frac{\omega}{2m}\right)}.
\end{align}
In this expression, we observe an exponential term with a correction factor. The exponential term corresponds to the usual commutative case with non-thermal radiation \cite{tunn5}, while the correction factor in our case of NC gauge theory depends on the Planck length. This factor introduces a quantum correction. In the semiclassical limit, where we set $l_{Planck}=0$, our expression reduces to the commutative one \cite{tunn5}.

Let us examine the correlation between successively emitted particles with different modes within the framework of NC gauge theory of gravity. Specifically, we will focus on the context of non-thermal radiation, starting with the first quantum emission carrying energy $\omega_1$. In this case, the emission rate \eqref{eq:tunneling3} takes the following form
\begin{align}\label{eq:tunneling4}
\ln\hat{\Gamma}_{\omega_1}=\frac{9\pi}{4}\Theta^2\ln\left(1-\frac{\omega_1}{m}\right)-8\pi m\omega_1\left(1-\frac{\omega_1}{2m}\right).
\end{align}
In accordance with Refs. \cite{correlation1, correlation2, correlation3}, the second quantum emission, characterized by energy $\omega_2$, is independent of the first emission, $\omega_1$. This independence holds true even in the presence of quantum gravity corrections \cite{correlation4}, and its emission rate is described by
\begin{align}\label{eq:tunneling5}
\ln\hat{\Gamma}_{\omega_2}&=\frac{9\pi}{4}\Theta^2\ln\left(1-\frac{\omega_2}{m}\right)-8\pi m\omega_2\left(1-\frac{\omega_2}{2m}\right).
\end{align}
The emission rate for two quantum emissions occurring simultaneously, with energies $\omega_1$ and $\omega_2$, is given by
\begin{align}\label{eq:tunneling6}
\ln\hat{\Gamma}_{\omega_1+\omega_2}&=\frac{9\pi}{4}\Theta^2\ln\left(1-\frac{(\omega_1+\omega_2)}{m}\right)\notag\\
&-8\pi m(\omega_1+\omega_2)\left(1-\frac{(\omega_1+\omega_2)}{2m}\right).
\end{align}

The statistical correlation between these events \cite{correlation1,tunn8,correlation4} is calculated using the expressions above, leading to
\small
\begin{align}\label{eq:correlation}
\hat{\chi}(\omega_1+\omega_2;\omega_1,\omega_2)&=\left(\frac{\ln\hat{\Gamma}_{\omega_1+\omega_2}}{\ln\hat{\Gamma}_{\omega_1}+\ln\hat{\Gamma}_{\omega_2}}\right)\notag \\
&=8\pi\omega_1\omega_2+\frac{9\pi}{4}\Theta^2\ln\left(\frac{m(m-(\omega_1+\omega_2))}{(m-\omega_1)(m-\omega_2)}\right)
\end{align}
\normalsize
Remarkably, the correlation function remains non-zero in the context of NC gauge theory of gravity, as well as in the commutative case (when $\Theta=0$, the commutative expression is recovered) \cite{correlation1, correlation2, tunn8, correlation3}. This implies that different radiation modes during black hole evaporation exhibit correlations with each other. Moreover, the presence of non-commutativity in this expression, enhances the statistical correlations between Hawking radiations. Furthermore, the presence of correlation between successively emitted quantum particles suggests that information can emerge within the Hawking radiation, thus addressing the information loss paradox. In this NC framework, information is preserved within this remnant, and the geometry enhances these correlations, enabling information to surface from the event horizon, akin to ``hidden messengers in Hawking radiation'' \cite{correlation1}.

\subsection{NC correction to the density number of particles emitted}\label{subsec:NCN}

As per Refs. \cite{number1, number2, number3}, the density number of particles emitted is intricately linked to the tunneling rate \cite{ntunn1, tunn4}. In our scenario, the density number resulting from the tunneling rate \eqref{eq:tunneling1} can be expressed as follows \cite{number1}
\begin{align}\label{eq:dn1}
\hat{n}=\frac{\hat{\Gamma}}{1-\hat{\Gamma}}=\frac{1}{e^{8\pi m\omega\left(1+\frac{3\Theta^2}{8m^2}\right)}-1},
\end{align}
Remarkably, the obtained expression takes the form of the Planck distribution of black body radiation and is intimately related to the NC parameter characterizing the deformed geometry. When $\Theta = 0$, this density number of particles emitted reduces to the commutative case \cite{hawking1}
\begin{equation}\label{eq:dn2}
\hat{n}=\frac{1}{e^{8\pi m\omega}-1}.
\end{equation}
\begin{figure*}[htb]
	\centering
	\includegraphics[clip=true,width=0.33\textwidth]{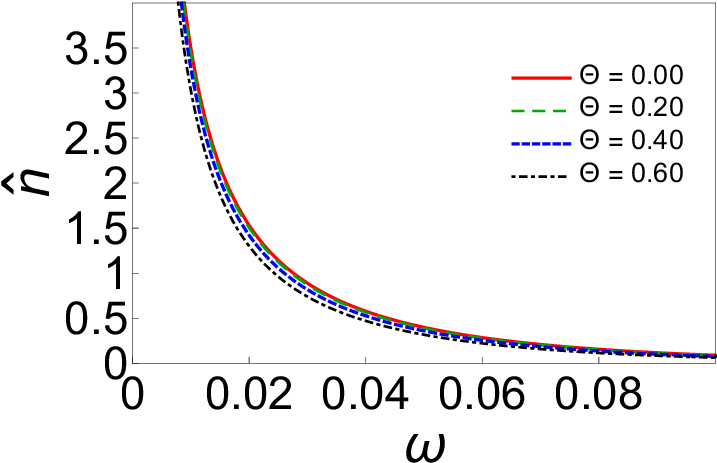}
	\includegraphics[clip=true,width=0.33\textwidth]{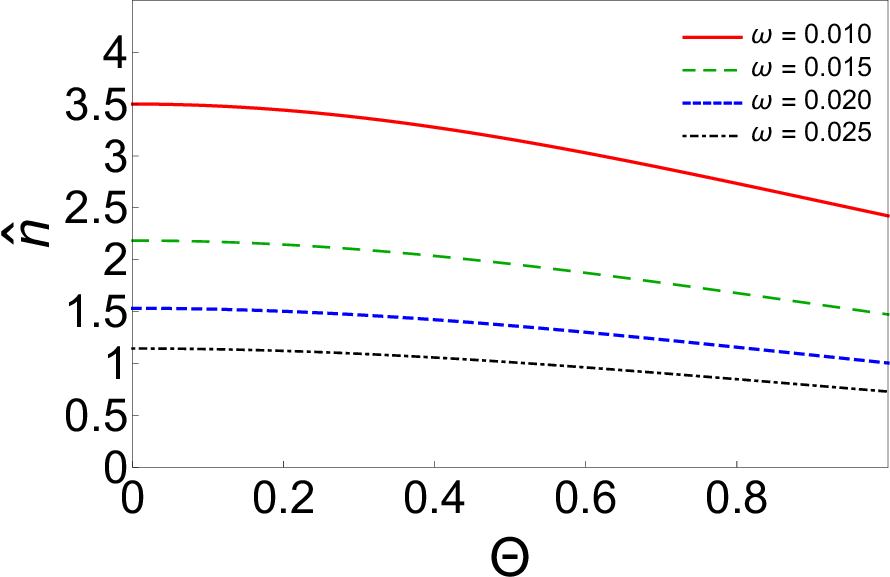}
	\includegraphics[clip=true,width=0.33\textwidth]{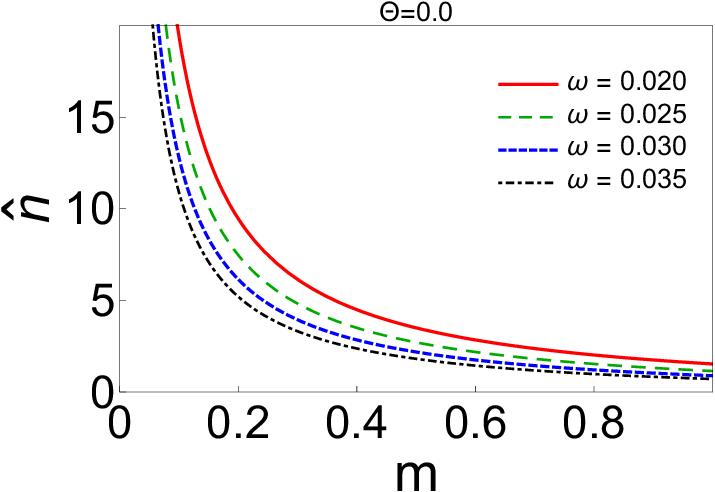}
	\caption{Density number $\hat{n}$ of particles emitted from a NC SBH as a function of particle frequency $\omega$ (left panel) with $m=1$ and different NC parameter $\Theta$. (Middle panel) density number of particles emitted as a function of the NC parameter $\Theta$ with $m=1$ and different frequency $\omega$. (Right panel) Commutative behavior $\Theta=0$ of $\hat{n}$ as a function of the black hole mass $m$, with different frequencies $\omega$.}
	\label{fig3}
\end{figure*}

The behavior of the density number of particles emitted through the tunneling process in the context of NC SBH as a function of particle frequency $\omega$ is illustrated in Fig. \ref{fig3} (left panel). A key observation in this figure is the influence of non-commutativity, where its effect results in a decrease in the density number of particles emitted by the black hole as the NC parameter $\Theta$ increases. This effect is analogous to the role of a potential well in quantum mechanics.

In the middle panel of Fig. \ref{fig3}, we demonstrate how non-commutativity decreases the density number of emitted particles, which is akin to the behavior observed in the commutative case where an increase in the black hole mass $m$ leads to a decrease in the density number of particles emitted (right panel). This indicates that non-commutativity plays a role similar to the black hole mass, effectively increasing the gravitational field of the black hole and explaining the reduction in the density number of particles escaping from the black hole.

\section{Conclusions}\label{sec:concl}

Our investigation within the framework of NC gauge gravity has revealed intriguing insights into the behavior of NC SBH. Firstly, we have constructed the deformed SBH employing a general form of tetrad fields. Then, we have extended the semi-classical tunneling approach \cite{tunn1, tunn2, tunn3} to the NC SBH, exploring two distinct scenarios that shed light on the interplay between non-commutativity, black hole properties, and quantum phenomena.

In the first scenario, we examined the case of tunneling particles from the NC SBH with massless characteristics and lower frequencies. This setting corresponds to pure thermal radiation, aligning with the expected behavior of Hawking radiation. Remarkably, we have established an equivalence between the quantum tunneling process and thermodynamic processes in the realm of NC spacetime. Our analysis of thermal radiation has unveiled the distinct impact of non-commutativity on the tunneling rate of these massless particles, akin to the role of a potential well in quantum mechanics. We have uncovered a remarkable result: the Hawking temperature reaches zero at a minimal radius $r_h^{min}\sim \Theta\sim l_p$ (for further details, see \cite{abdellah2}), where non-commutativity halts the complete evaporation of the black hole, resulting in a remnant. Furthermore, our estimation of the NC parameter $\Theta$ suggests that spacetime non-commutativity becomes manifest at the Planck scale.

In the second scenario, we introduced the conservation of energy \cite{tunn2}, particularly in the context of large frequency emissions. This approach yielded a deviation from thermal radiation in the tunneling rate, consistently aligning with an underlying unitary quantum theory where the tunneling rate is related to changes in black hole entropy. Our findings have unveiled a logarithmic correction as the NC correction to the entropy of NC SBH. This correction is consistent with other quantum gravity theories, including String Theory (ST), Quantum Loop Gravity (QLG), Generalized Uncertainty Principle (GUP), and Modified Dispersion Relation (MDR) \cite{QLG1, QLG2, ST3, nozari3, nozari4}. Moreover, the coefficient of this logarithmic correction within NC gauge theory of gravity is directly related to the Planck length $l_{Planck}$, representing a quantum surface at the Planck scale.

We then investigated the correlation between successively emitted particles with different frequencies $\omega$ in NC spacetime for a non-thermal emission. The results have highlighted the presence of correlations between successive emissions, with non-commutativity playing a role in enhancing these correlations within Hawking radiation. This enhancement not only preserves information within the remnant black hole but also provides a compelling solution to the information loss paradox.

Finally, we explored the NC correction to the density number of particles created by the NC SBH in the scenario of pure thermal radiation at a lower frequency approximation. Our results confirm the effect of non-commutativity, leading to a decrease in the density number of particles.

\acknowledgments
This work is supported by PRFU Research Project B00L02UN050120230003, Univ. Batna 1, Algeria.

\bibliography{ref}

\end{document}